\documentstyle[12pt,epsfig]{article}
\date{}
\title{$\gamma$-quanta propagation in single crystals}  
\author{V.A.Maisheev \thanks{E-mail maisheev@mx.ihep.su} \\
{\it Institute for High Energy Physics, 142284, Protvino, Russia }}

\begin{document}
\large
\maketitle
\def\arcctg{\mathop{\rm arcctg}\nolimits} 
\def\ch{\mathop{\rm ch}\nolimits}
\def\sh{\mathop{\rm sh}\nolimits} 
\normalsize
\begin{abstract}
  Propagation of $\gamma$-quanta in the single crystals, oriented in a region 
of the coherent pair production is considered. The qualitative description
of the process is also discussed. The theory of $\gamma$-quanta propagation
in the anisotropic medium is illustrated by the help of the particular 
calculations of such a process in silicon single crystals.
 
  It is shown that the single crystals are sensitive to the initial circular
 polarization of  $\gamma$-beam despite the fact that the cross section
of absorption is independent of it. The reason  is that the normal 
electromagnetic waves (an eigenfunctions of the problem) are elliptically
polarized. The speeds of absorption and motion of both the normal waves are 
different and as a result the process of $\gamma$-quanta propagation depend 
on the initial polarization state. The calculated value of asymmetry is 
about 20\%  for 25 GeV $\gamma$-quanta propagating in 100 cm silicon   
single crystal.

   The obtained results are useful in creating of polarimeters for
high energy electrons and $\gamma$-quanta.

\end{abstract}
\section{ Introduction}
It is well known that the optical properties of a medium may be described 
with the use of a permittivity tensor \cite{LL_ME,AG}. The anisotropic
medium referred as such a medium in which the permittivity tensor is 
a symmetric tensor \cite{LL_ME,AG}. In the general case the components of
this tensor are complex values and, because of this,  the complex permittivity
tensor does not reduce to principal axes, except in specific cases.  
The simplest example of such an anisotropic medium of the general type is 
a combination of the two linearly polarized monochromatic laser waves with 
different frequencies (dichromatic wave) moving in the same direction. 
In general the directions of linear polarization of these monochromatic
waves are different. Another more complicated example is a single
crystal oriented in the region of coherent $e^-e^+$-pair production 
\cite{U,Ter,C}. 

The permittivity tensor was found for both these examples in 
papers\cite{MVA,MMF}.
Besides, in \cite{MVA}  the theory of $\gamma$-quanta propagation in the 
anisotropic medium of a general type is also considered. 
 One of the main results of these papers is the prediction of sensitivity of the 
anisotropic medium to the initial circular polarization of propagating 
$\gamma$-quanta. It is important for creating of devices measuring the 
circular polarization of $\gamma$-quanta or longitudinal polarization of 
electrons. In the last case the electron beam is transformed into 
bremsstrahlung $\gamma$-beam by the use of amorphous target \cite{ROA}. 
The circular
polarization of the end of bremsstrahlung spectrum is approximately equal to
the longitudinal polarization of a primary electron beam.
    In this paper we consider the theoretical basis for creating of the
 $\gamma$-beam polarimeters. 

\section{Qualitative consideration of process}
 As already noted, the  theoretical description of $\gamma$-quanta 
propagation in the anisotropic medium is contained in paper \cite{MVA}. 
The permittivity tensor in single crystals oriented in the region of 
coherent $e^-e^+$-pair production is found in \cite{MMF}. In the case of
high energy $\gamma$-quanta the process of propagation is determined
primarily by the transverse part of the permittivity tensor, while
the longitudinal components of the tensor are higher-order infinitesimals
in the interaction constant $\alpha$ \cite{BT,BKF}. In this way 
the permittivity tensor in anisotropic medium is a symmetric tensor 
of rank two.    

Now we
make an attempt of the qualitative consideration of the $\gamma$-beam
propagation in the anisotropic medium.  
 As is well known that the imaginary components of permittivity tensor in
anisotropic medium describe the absorption of $\gamma$-quanta. The cross
section $\sigma_{e^-e^+}$ depends on the linear polarization of $\gamma$-
quanta
\begin{equation}
 \sigma_{e^-e^+} = \cal{A} +\cal{B}(\bf{e\tau})^2,
\label{1}
\end{equation}
where the unit vectors $\bf{e}$ and $\bf{\tau}$ determine the polarization plane
of the $\gamma$-quanta and some definite plane in single crystal (the wave 
vector of the $\gamma$-quanta lies in both planes). The formula (1) 
essentially determines a symmetric tensor of rank two, whose components 
we denote as $\sigma_{kl}\,(k,l=1,2) $.  Then the imaginary parts of
the permittivity tensor components are 
\begin{equation}
\varepsilon^"_{kl} =  {{N\sigma_{kl}c } \over{ \omega}} 
\label{2}
\end{equation}
where N is the atomic density of single crystal, c is the speed of light  
and $\omega$ is the $\gamma$-quanta frequency.
 The real components of the permittivity tensor we can find with the help
of the despersion relations \cite{AG}
\begin{equation}
\varepsilon'_{ij}- \delta_{ij}= {2\over \pi} 
{\cal P}\int_{0}^{\infty} {{x\varepsilon''_{ij}(x)\, dx}
\over {x^2-\omega^2}}
\label{3} \, ,
\end{equation}
As is seen from (\ref{1}) the cross sections for $\gamma$-quanta with
direction of linear polarization  parallel and perpendicular with respect to
unit vector $\tau$ are different. As is well known that the ellipse is
a geometric interpretation of the symmetric tensor. These ellipses are
represented on Fig.1 for three  energies of $\gamma$-quanta. Let take
the above-mentioned dichromatic laser wave as an example of the anisotropic 
medium. Fig.1a illustrates the behavior of the absorption ellipses at 
different energies in case when linear polarizations of the both waves are
parallel or perpendicular with respect to unit vector $\bf{\tau}$.
Fig.1b illustrates the similar behavior, but the angle between polarizations
of the two waves is not equal to 0 or $\pi/2$. In both cases the cross
sections are described by the equation (\ref{1}). However, 
in the case on Fig.1a
the common symmetric plane exists for all energies of $\gamma$-quanta.  
This is true because of the natural symmetry of the problem. 
As Fig.1.b illustrates, in this case the symmetric plane also exists
but the position of the plane in space is different  for every energy
of $\gamma$-quanta. In particular the position of this plane depends on
the ratio of intensities of the dichromatic wave components. It is obviously
that the symmetry plane is determined in main by the strong wave
when the intensity one of two waves is very weak in the comparison with
another. In such a manner, in this case the symmetry have the dynamic
character, which defined by the particular  mechanism of the interaction.   
The numerical quantities  of this interaction are different for every
energy of $\gamma$-quanta. Hence it follows that dynamic symmetry plane
rotate at changing of the $\gamma$-quanta energy.
   
  We can see from Eq(\ref{3}) that real components
of permittivity tensor is determined by  the integration over all energies
of $\gamma$-quanta. It is means that in general the position  of the
symmetry plane for real permittivity tensor differ from the similar position
for imaginary part of this tensor. In other words, the axes of both tensors
are not parallel or perpendicular in between. 
  Hence we get that in general the preferred two planes exist in a space 
for propagating $\gamma$-quanta of a fixed energy.
 
As is well known that $\gamma$-beam in a medium present the superposition
of two states. These states referred as normal electromagnetic waves \cite{AG}.
The normal waves are the eigenfunctions of the problem and they have 
determinate polarization characteristics, which depend on the optical
properties of a medium. So, the $\gamma$-quanta, having the polarization
characteristics identical to one of normal waves, are conserve their at the
propagation in a medium. In general the speeds of absorption of normal waves 
are different.

 It is evidently that both normal waves are linearly polarized in case 
illustrated in Fig.1a. These polarizations are perpendicular in between
and their directions coincide with the principal axes of ellipses.
In other case (see Fig.1b), in general the principal axes of both 
ellipses are not parallel, and it is believed that normal wave are
elliptically polarized. It is also follows from paper \cite{MMF}.  
In the case, when the absorption in medium is absent, the all imaginary
components of permittivity tensor are equal to zero and because of this
the normal waves are always linearly polarized.  

 In the general case the initial  polarization state of $\gamma$-quanta 
changes at  propagation in a medium. The intensity of $\gamma$-beam
on thickness x in the anisotropic medium can calculate by the help of 
follows relation \cite{MVA}
\begin{equation}  
 J_{\gamma}(x)= A(x) +B(x)\xi_1 +C(x)\xi_2 +D(x)\xi_3
\label{4}
\end{equation}
where $ \xi_i,\, (i=1,3)$ are the initial Stokes parameters of $\gamma$-quanta, 
A,B,C,D are simple functions of x including also some parameters as, for 
example, refractive indices of normal waves. When the normal waves have
only linear polarization the function C(x) is equal to zero on any
thickness x. In the case, when the normal waves have nonzero component of
circular polarization the function C(x) is equal to zero at x = 0 and in 
general it is nonzero at $x>0$. Besides, the first derivative of the
C(x)-function is equal to zero at $x=0$. It means that the cross section
of  the absorption process independent of the circular polarization of
the $\gamma$-quanta. Here we employ the well-known 
formula $dJ= - J(0) N \sigma dx$ for thin targets.
 
In this way the relation (\ref{4}) show that intensity of $\gamma$-quanta
propagating in the anisotro\-pic medium depends on the initial polarization
despite the fact that the cross section of absorption is independent of it.
The reason is that the normal waves are elliptically polarized. The speeds
of absorption and motion of both the normal waves are different due to
the process $\gamma$-quanta propagation depends on the initial polarization
state.
On the other hand the initially unpolarized $\gamma$-beam became
elliptically polarized \cite{MVA}.
 This is an essential prerequisite for creating of the $\gamma$- polarimeter.
To do this requires computations of the $\gamma$-quanta propagation in
single crystals.

\section{$\gamma$-quanta propagation in single crystals}
 In this section we present some results of the calculations of
$\gamma$-beam propagation in silicon single crystals.
 The investigated process is determined with the help of such parameters
as refractive indices and polarization states of normal electromagnetic 
waves. The imaginary parts of refractive indices are responsible for
the speed of normal wave absorption. The real parts and 
polarization states of normal waves  are also in term C(x) 
(see equation (\ref{4})). Let us define the polarization state of one 
normal wave with the help of Stokes parameters $X_1,\, X_2, \, X_3
(X_1^2+X_2^2+X_3^2=1) $.
Then the polarization state of another wave is correspondingly equal
to $ -X_1,\, X_2,\, -X_3$. Intuition suggest that the parameter $X_2$
should be significant for  circular polarization detection. 
As it is follows from \cite{MMF} parameter $ X_2$ is significant when 
$\gamma$-beam moving under no large angle with respect to any strong
crystallographic axis. We select the $<001>$ axis in silicon for calculations.
We also select the Cartesian coordinate system with one axis along
$<001>$ axis and two another axes along $<110>$ and $<1-10>$ axes. Direction
of $\gamma$-quanta motion we can determine with the use of angle
$\theta$ with respect to $<001>$ axis and azimuth angle $\alpha$
around of this axis ($\alpha =0$, when the $\gamma$-quanta momentum 
lies in the $(110)$-plane).  However another angles are more convenient
to use $\varphi_H = \theta \cos{\alpha}, \varphi_V=\theta \sin{\alpha}$.

The characteristics of process have the same numerical values at  
different $\gamma$-quanta energies $E_\gamma$ when  the following invariant 
parameters are used: $W_H=E_\gamma G_2 \varphi_H/(2mc^2),$  
$W_V= E_\gamma G_3 \varphi_V/(2mc^2)$. Here $m$ is the electron mass,
$G_2, G_3$ are the reciprocal lattice constants (in our case $G_2= G_3 
= 0.01264$ in units of the inverse electron Compton length).
In the case being considered we can write these parameters as
$W_H=12.366 E_\gamma \varphi_H, W_V =12.366E_\gamma \varphi_V [GeV radian] $. 
The calculations are carried out for $E_\gamma =25 GeV$. 
The Figs.2-6 illustrate these results of calculations. In the calculations
the Moliere form factor was employed \cite{BKF}. 
 
As Fig.3 illustrates,
the regions with high circular polarization $X_2$ exist in single crystals.
The absolute value of $X_2$ is near 1 at $\varphi_H=3.008$ mrad and
$\varphi_V=3.558$ mrad. We can see that the  high quantity of $X_2$ 
is observed in region $\approx \pm 0.1$ with respect to central $\varphi_H$,
$\varphi_V$-values. The direction, in which the circular polarization $X_2$
is equal to 1, is the direction of the so-called singular axis,
described in the crystal optic of the visible light \cite{AG}.  
We can see in Fig.2 that refractive indices are approximately  equal to one
another near this direction.
Notice, that the value $Im(n)-\varepsilon_A/2$, as a function of invariant
$W_H,W_V$-parameters, is independent of the $\gamma$-quanta energy. The
$\varepsilon_A$-value  is determined in \cite{MMF}, and for the silicon   
single crystal is numerically equal to $1.32\, 10^{-15}/E_{\gamma}$.   
Fig.5 illustrates the variation of polarization state of the  propagating 
$\gamma$-quanta. We can see that the initially unpolarized beam obtain
in general the linear and circular polarization. Notice, that equations
of paper \cite{MVA} break down for $ X_2=1$ (see also \cite{AG}). However 
they are true in the neighborhood of this point and our calculations are made
for $ X_2 =0.995$. 

Figs.4 and 6 illustrate the absolute and relative losses of $\gamma$-quanta
intensity. One can see that these losses  depend on the initial
polarization of the $\gamma$-quanta. We select the value $A_s = |I_p-I_0|/I_0$
as the degree of relative losses of intensity, where $I_0$ is the losses
of initially unpolarized beam and $I_p$ is the losses of completely circularly
polarized beam. 

Notice, that the initial  polarization of $\gamma$-quanta 
is not needed to change for measuring of the asymmetry $ A_s$. So, let 
$\gamma$-beam is completely circularly polarized at the entry of the
single crystal which have the orientation angles are equal to $\varphi_{H1}$ 
and $\varphi_{V1}$. 
After measuring the intensity losses  we change the single
crystal orientation so, that new angles are equal to $\varphi_{H2}=\varphi_{V1}$
and $\varphi_{V2}=\varphi_{H1}$. Then we get that
$ |I_1-I_2|/0.5(I_1+I_2)=|I_1-I_2|/I_0=2A_s$.     

 As already noted, the symmetry of problem have in general the dynamic
character. It is means that  our knowledge of such characteristics as
refractive indices and $X_i$-values depends on the possible of
uncertainty in the initial data for calculations. We think that  
main uncertainty is connected with the electric fields of single crystals.
We carry out calculations of $X_2$-value with the use of experimental
silicon form factors\cite{Si,MV}, instead Moliere ones. The results of these
calculations agree closely with each other. However, a little
displacement of the singular axis direction (when $X_2=1$) take a place 
(in this case $W_H=0.975, W_V=1.075$, instead $W_H=0.93,W_V=1.10$ for
Moliere potential).  

\section{Discussion and Summary}
The results of calculations in the silicon single crystal oriented near  
$<001>$ axis show detectable sensitivity to the circular polarization of  
propagating $\gamma$-quanta. Below we discuss some questions of the
limitations and optimization.

 Our consideration of the process are based on the traditional theory
of the coherent $e^+e^-$-pair production in single crystals \cite{U,Ter}. 
However, this process is transformed into the analogous process in  
a "strong field" of the crystallographic axis \cite{RMo} and the mathematical
description of the both processes is different. The process in the
"strong field" take a place at angles $\phi_B \le V_a/mc^2$ with respect to
axis, where $V_a$ is the potential of this axis. 
In the case of $<001>$ silicon axis this angle is equal to $\approx 0.15$ mrad.  
It is means that the upper bound for our consideration ($<001>$,Si) 
is $\sim 500 GeV$. 
 
We believe that the asymmetry of process $A_s$ on the 
same thickness  is growing with the $\gamma$-quanta energy increasing.
It is true because of the increasing of the cross section of the
$\gamma$-beam absorption. 
As Fig.6 illustrates, the detectable value $A_s$ take a place for 
orientations with not high quantity of the $X_2$-value. 
This orientation region is more wide and flat then region near singular
axis. It is important for measuring polarization of the $\gamma$-beam
with a large angle divergence. 

It is well-known that the $<011>$ and $<111>$ axes in silicon single crystals
have more  strong electric fields, then $<100>$ axis. Because of this, 
it is expected more high $A_s$-values for these axes. 

 We believe that the theoretical prerequisites exist for creating 
$\gamma,e$-polarimeters on the base of the considered here phenomenon.
However, a lot of optimization calculations need to be done before for it.

\newpage

\begin{figure}[h]
\begin{center}
\parbox[c]{14.5cm}{\epsfig{file=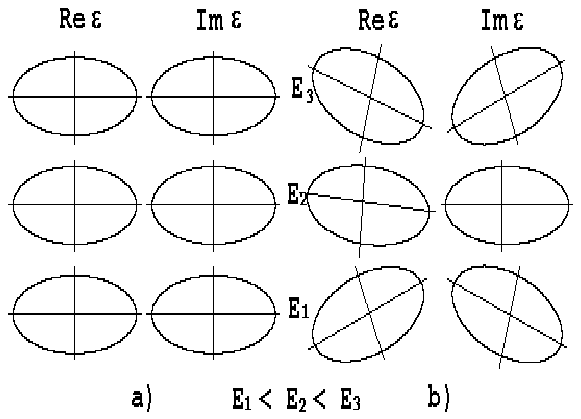,height=10cm}}
\parbox[c]{15cm}{\caption{
 The geometrical interpretation of permittivity tensor in anisotropic              
 medium. Further explanations are given in the text.
              }}  
\end{center} 
\end{figure}

\begin{figure}
\begin{center}
\parbox[c]{12.5 cm}{\epsfig{file=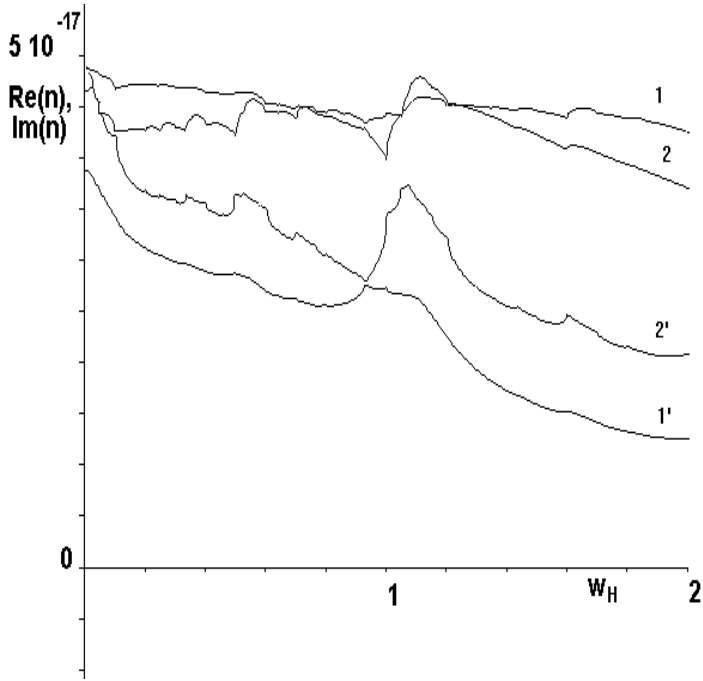, width=10 cm}}
\parbox[c]{15cm}{\caption{
  Real minus unit $(1', 2') $ and imaginary $(1,2)$ parts of the refractive 
  indices near  of $<001>$-axis  in silicon single crystal as a function of 
  the parameter $W_H$. $E_{\gamma} =25 GeV,\,W_V= 1.10$. Temperature of  
  single crystal $T_{Si}=300^\circ$ K. 
              }}  
\end{center}
\end{figure}

\begin{figure}
\begin{center}
\parbox[c]{12.5 cm}{\epsfig{file=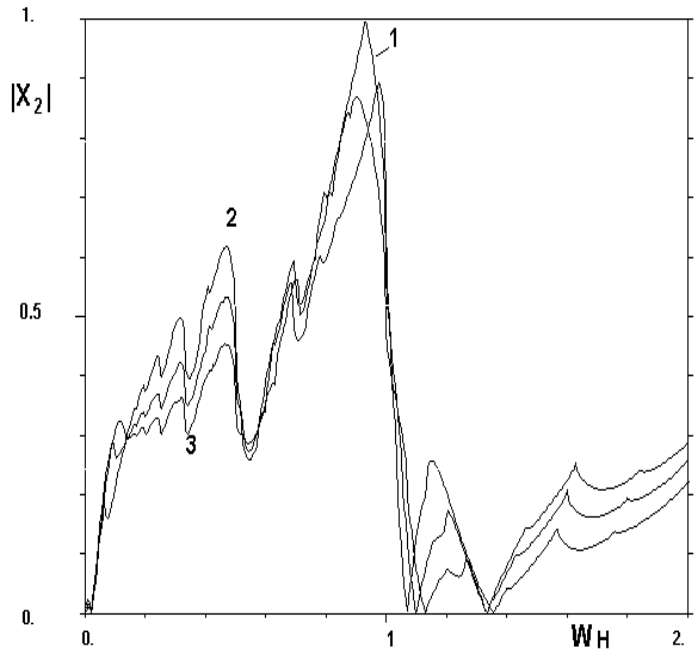, width=10 cm}}
\parbox[c]{15cm}{\caption{
Absolute quantity of normal wave circular polarization $X_2$  near 
  of $<001>$-axis  in silicon single crystal as a function of the 
  parameter $W_H$. $W_V$ = 1.10, 1.13, 1.07 for 1,2,3-curves.
  \,$T_{Si}=300^\circ$ K.   
              }}  
\end{center}
\end{figure}
\begin{figure}
\begin{center}
\parbox[c]{12.5 cm}{\epsfig{file=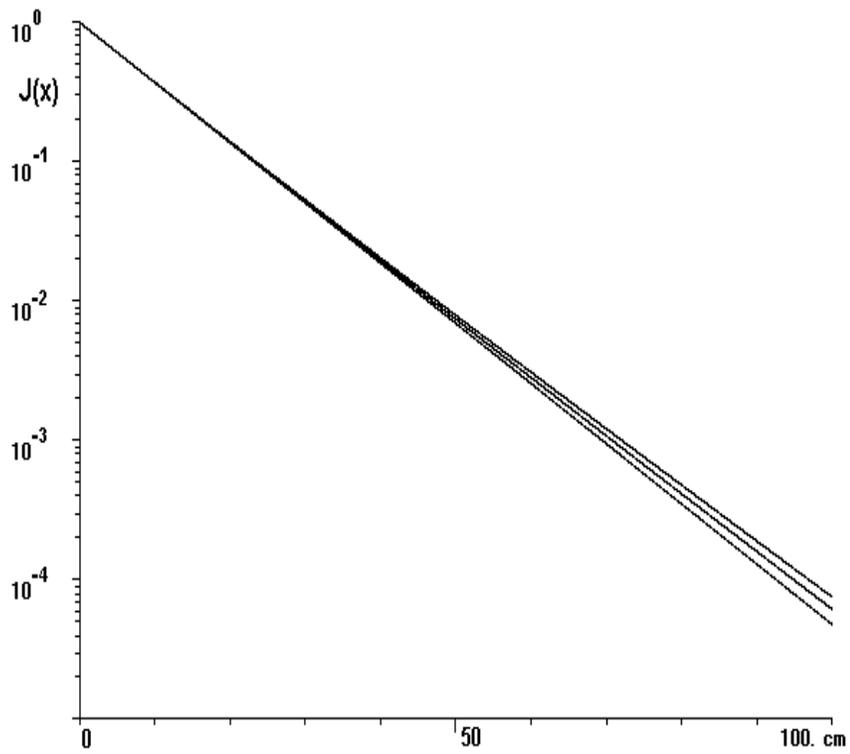, width=12 cm}}
\parbox[c]{15cm}{\caption{
  Intensity of 25 GeV $\gamma$-quanta as a function of the 
  single crystal thickness. Central curve is the intensity of
  initially unpolarized beam, upper and bottom curves are the
  intensities of beam with the corresponding initial circular  polarization
  equal to -1 and +1. $W_H=0.93,\,  W_V= 1.10$, $T_{Si}=300^\circ$ K.  
              }}  
\end{center}
\end{figure}
\begin{figure}
\begin{center}
\parbox[c]{12.5 cm}{\epsfig{file=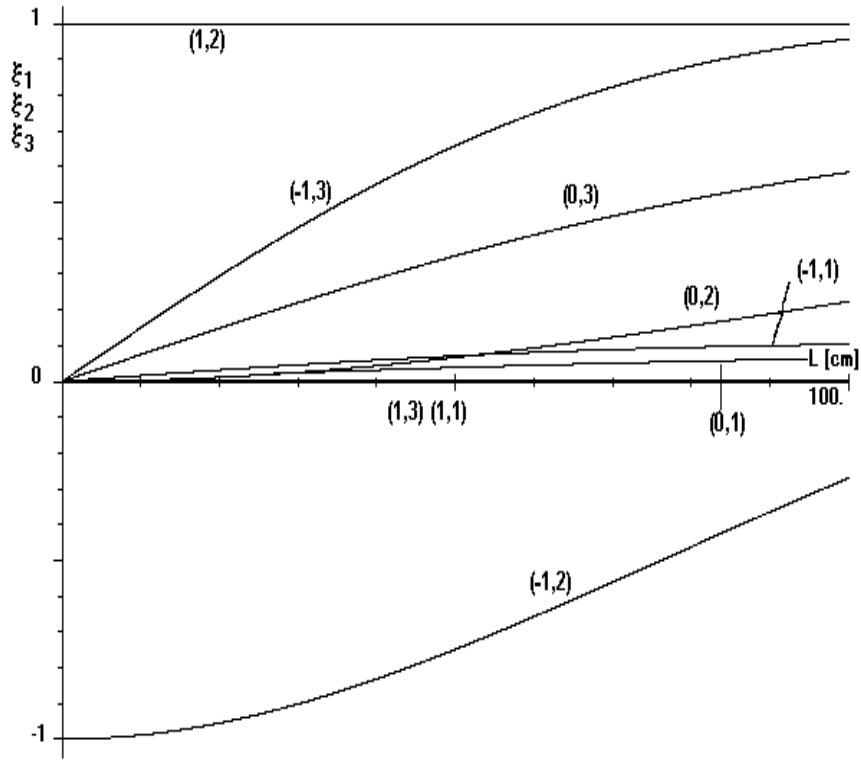, width=12 cm}}
\parbox[c]{15cm}{\caption{
  Stokes parameter variations of 25 GeV $\gamma$-quanta as a function of the 
  single crystal thickness. The first number in the parenthesis is initial
  circular polarization (1,0,-1), the second number is number of Stokes
  parameter i=1-3. $W_H=0.93,\,  W_V= 1.10$, $T_{Si}=300^\circ$ K.
              }}  
\end{center}
\end{figure}
\begin{figure}
\begin{center}
\parbox[c]{12.5 cm}{\epsfig{file=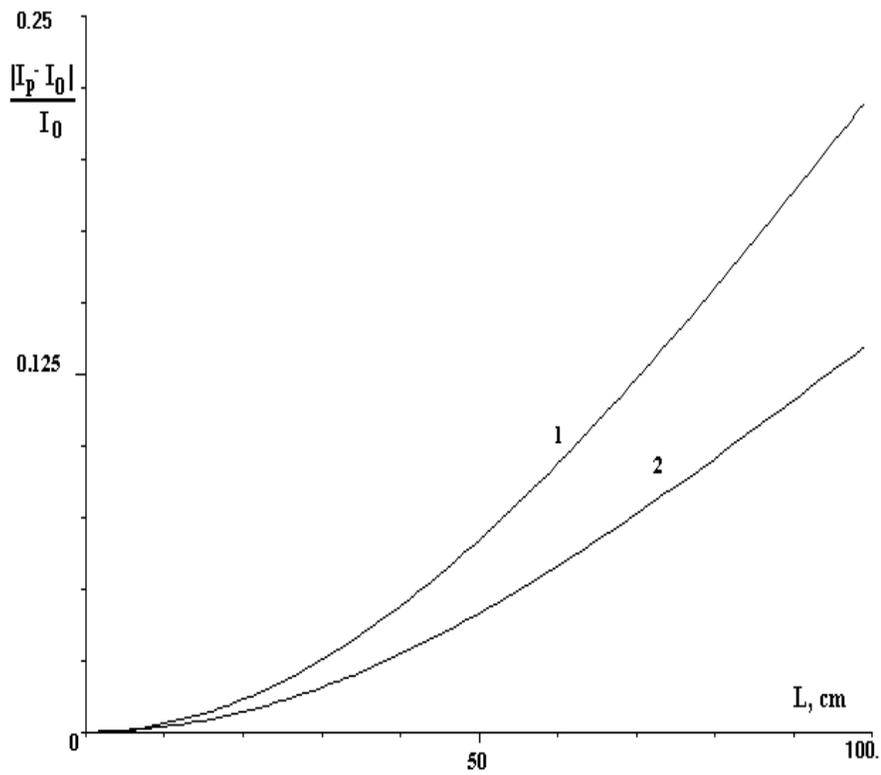,width=12 cm}}
\parbox[c]{15cm}{\caption{
 Asymmetry of the process as a function of the silicon single crystal             
 thickness. Curve 1 is for $W_H=0.93,\, W_V=1.10$ and curve 2 is for 
 $W_H=2.0,\, W_V=1.10$.$E_\gamma =25 GeV$, $T_{Si}=300^\circ$ K.
              }}  
\end{center}
\end{figure}

\end{document}